\documentclass{ws06-procs}

\usepackage{graphicx}

\begin{document}

\title{Dark Matter of the Universe}

\author{Azzurra Auteri}

\address{Dipartimento di Scienze Fisiche, Universit\`{a} di Napoli ''Federico II'', Complesso Universitario Monte Sant'Angelo, Via Cinthia,1 - 80126 Napoli, Italy}

\maketitle

\abstracts{
Introduction to the meaning of dark matter and a few historical gestures. What it is? By what the dark matter is formed? Experimental phenomenon evidences and relative explanation under the point of view of the physics of the elementary particles. Conclusions with gestures at a few studies who at present are responsible for this subject.
}

Talk gave at Workshop was titled ``DARK MATTER OF UNIVERSE''.
It's  a very simply introduction of this study, with help of some picture that 
describe exsperimental results I have illustrateded the main Dark Matter's 
phenomena.
At the end there's a list about more important past and presernt experiment  
of this study and a little bibliography.

I  have discussed about six subject:

\begin{itemize}
	\item What's the dark matter?
	\item First experimental evidence
	\item From what is maked up dark matter?
	\item How much dark matter is in the universe?
	\item Dark energy
	\item Experiment
\end{itemize}

\begin{figure}[ht]
  \begin{center}
    \includegraphics[width=10cm]{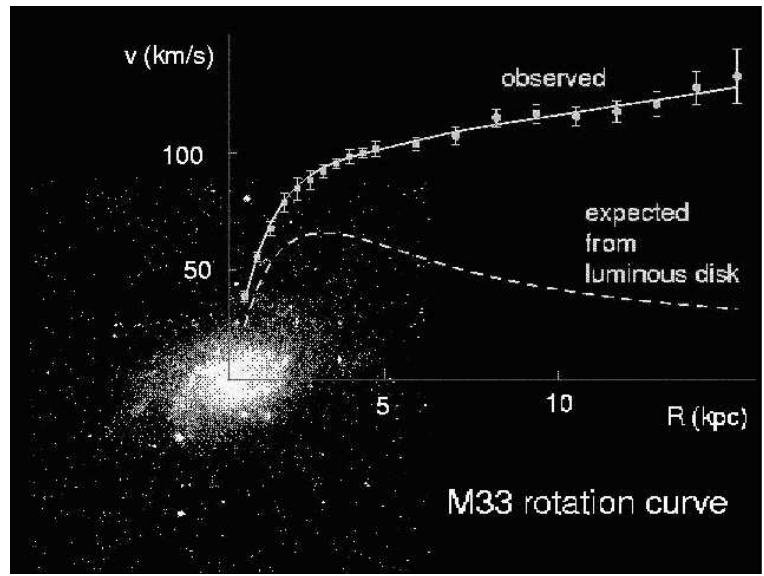}
  \end{center}
\caption{The observed course indicates that the tangential speed of a star grows to increasing of the distance from the center and that indicates the presence of one great amount of invisibile matter. \label{foto}}
\end{figure}

At the beginning I introduce elementary definition about dark matter due at observation of universe electromagnetic field.
Thanks Zwicky and Smith's work it was noted down an upper speed inside galaxy that General Gravitation didn't foresee. It was brought to an end that in the universe there is a deficit mass called ``DARK MATTER''  because it isn't directly observable but it's deduced for example by star motion inside a galaxy.
The figure \ref{foto} is a prove of the existence of the dark matter.

What makes up dark matter is not simple to say.
The scientists are sure that it must have characteristic following:

\begin{itemize}
	\item It exists from various billions of years and therefore the constituted particle must be stable;
	\item The dark matter is concentrated around to the galactic nuclei us and must be able to become a member of;
	\item It isn't luminous matter and therefore does not interact with the electromagnetic radiation if not in negligible way.
\end{itemize}

The particle that have this characteristc at the same time are : neutrinos, assions and wimps.
The neutrinos are particle producted by nuclear radiation or radiactive decay.
The assions are particle expected by strong interaction theory and can make cosmic matter.

The wimps is an acronym for ``Weak Interactive Massive Particle''  particles expected by super-symmetry.
For the ammount of dark matter introduce a parameter caller ``OMEGA'', it's a ratio between density of universe matter and critical density. With OMEGA we   define space geometry and can estimate  Universe expansion. Trough this study is possible maintain that the Universe is maked up for 99,5\%  by Dark Matter.

To justify the accelerate Universe expansion we introduce the ``DARK ENERGY'' , the vacuum energy. This is confirmed thank to background cosmic radiation.

The study of Dark Matter can be developed with two way : an direct measure and Indirect measure. In the first case we have to study interaction between particle inside a detector (called experiment with accelerator), in the second case we have to study for exsample , neutrinos  produced for annilation inside the sun (called experiment without accelerator) .

MACRO, ANTARES and  AMS are experiments without accelerator instead  L3. DELPHY, OPAL, CDF and CMS using the accelerator.


\begin{thebibliography}{0}
\bibitem{ja} Bonomett S., {\it Materia Oscura in Enciclopedia della Scienze Fisiche,vol.III}
   (Istituto dell'Enciclopedia Italiana, Roma) (1994), pag. 442-445.


\bibitem{ppz} Scaramella R., {\it Stuttura su grande scala dell'Universo in Enciclopedia della Scienze Fisiche, vol.VI}     (Istituto dell'Enciclopedia Italiana, Roma) (1995), pag. 436-445.

\bibitem{ma} Nell'esperimento Italiano DAMA  ricerca diretta dei Wimps:
          http://www.lngs.infn.it/lngs/htexts/dama/.


\bibitem{bu} Bond J. R., Efstathiou G., Silk J., {\it Phys. Rev. Lett.} {\bf 45} (1980).

\bibitem{bd} Peebles P. J. E., {\it Principles of Physical Cosmology} (Princeton UniversityPress, Princeton NJ) (1993)

\end{thebibliography}
\end{document}